\documentclass[11pt]{book}
\usepackage[a4paper,vmargin=3cm,hmargin=3.1cm,headheight=13.1pt]{geometry}

\usepackage{amsmath,amsfonts,graphicx,subfigure}
\usepackage[makeindex]{imakeidx}
\usepackage{color,hyperref}

\newcommand{\chapIndykA}{8}

\newcommand{\chapPolyhedralMaps}{20}

\newcommand{\chapShortest}{31}

\newcommand{\chapGraphDraw}{55}

\newcommand{\gglos}[1]{\index[auth]{#1}}
\newcommand{\tindex}[1]{\index[term]{#1}}

\usepackage{pbox,bm,titlesec}
\usepackage{marvosym}
\usepackage{mathrsfs}
\usepackage[psfrag]{xfigfrag} 

\long\def\ghost#1{\begin{picture}(0,0)\put(0,0){#1}\end{picture}}

\newsavebox{\myvcenter}\newlength{\myvcenterheight}
\newcommand{\vcent}[1]{\settoheight{\myvcenterheight}{#1}%
  \addtolength{\myvcenterheight}{-.5\baselineskip}%
  \raisebox{-.5\myvcenterheight}{#1}}

\newcommand{\surf}{S}
\newcommand{\FF}{\mathscr{F}}

\newcommand{\Z}{\ensuremath{\mathbb{Z}}}
\newcommand{\R}{\ensuremath{\mathbb{R}}}
\newcommand{\SSS}{\ensuremath{\mathbb{S}}}

\newcommand{\E}{\ensuremath{\operatornamewithlimits{\mathbb{E}}}}

\newcommand{\inv}{\ensuremath{^{-1}}}

\newcommand{\cut}{{\setminus\mskip-6.5mu\setminus}}

\newcommand{\pf}[1]{\psfrag{#1}{$#1$}}

\newcommand{\poly}{\ensuremath{\text{poly}}}

\newcommand{\partabular}[4]{%
 \ensuremath{\left#1\begin{tabular}{#3}#4\end{tabular}\right#2}%
}

\usepackage{hyperref,fancyhdr}

\titleformat{\section}[block]{\fontsize{14}{16}\bfseries}{}{0pt}{%
\begin{minipage}{\textwidth}
\hrule height .5pt width \textwidth\kern 1mm
\hrule height .5pt width \textwidth
\end{minipage}\\\thesection\ \ }[\vspace{2pt}]

\titleformat{\subsection}[block]{\fontsize{12}{14}\bfseries}{}{0pt}{%
\begin{minipage}{\textwidth}
\hrule height .5pt width \textwidth
\end{minipage}\\}[\vspace{1.2pt}]

\titleformat{\subsubsection}[block]{\fontsize{11}{13}\bfseries}{}{0pt}{\\}[\vspace{.8pt}]

\def\A#1{\section{#1}}
\def\Bnn#1{\subsection*{#1}}
\def\C#1{\subsubsection*{#1}}

\newenvironment{gllist}{\begin{itemize}}{\end{itemize}}
\def\trmbitx{\bf\em}

\makeatletter
\renewenvironment{thebibliography}[1]
     {\subsection*{REFERENCES}%
      \list{\@biblabel{\@arabic\c@enumiv}}%
           {\settowidth\labelwidth{\@biblabel{#1}}%
            \leftmargin\labelwidth
            \advance\leftmargin\labelsep
            \@openbib@code
            \usecounter{enumiv}%
            \let\p@enumiv\@empty
            \renewcommand\theenumiv{\@arabic\c@enumiv}}%
      \sloppy
      \clubpenalty4000
      \@clubpenalty \clubpenalty
      \widowpenalty4000%
      \sfcode`\.\@m}
     {\def\@noitemerr
       {\@latex@warning{Empty `thebibliography' environment}}%
      \endlist}
\makeatother

\makeatletter
\def\blfootnote{\gdef\@thefnmark{}\@footnotetext}
\makeatother
\begin{document}
\setcounter{chapter}{23}
\setcounter{section}{0}
\setcounter{page}{1}

\pagestyle{fancy}

\fancyhf{}
\lhead[\small\textit{\textsf{\thepage\quad \'E.~Colin de Verdi\`ere}}]{}
\rhead[]{\small\textit{\textsf{Chapter 23: Computational topology of graphs on surfaces\quad\thepage}}}

\thispagestyle{empty}

\vspace{-1pc}

\noindent{\sf\LARGE 23\quad COMPUTATIONAL TOPOLOGY}

\vspace{7pt}

\noindent{\sf\LARGE \hphantom{23}\quad OF GRAPHS ON SURFACES}

\vspace{1pc}

\noindent{\sf\LARGE \hphantom{23}\quad \Large \'Eric Colin de Verdi\`ere}

\vspace{4pc}

\blfootnote{Author's affiliation: CNRS, LIGM, Universit\'e Paris-Est
  Marne-la-Vall\'ee, France. Email:
  \texttt{eric.colindeverdiere}\texttt{@u-pem.fr}. Part of this work was
  done while the author was at CNRS, D\'epartement d'Informatique, \'Ecole
  normale sup\'erieure, Paris, France.}

\Bnn{INTRODUCTION}

\noindent
This chapter surveys computational topology results in the special,
low-dimensional case where the ambient space is a surface.  Surface
topology is very well-understood and comparably simpler than the
higher-dimensional counterparts; many computational problems that are
undecidable in general (e.g., homotopy questions) can be solved efficiently
on surfaces.  This leads to a distinct flavor of computational topology and
to dedicated techniques for revisiting topological problems on surfaces
from a computational viewpoint.

Topological surfaces and graphs drawn on them appear in various fields of
mathematics and computer science, and these aspects are not surveyed here:
\begin{itemize}
\item in \emph{topology} of three-dimensional manifolds, also in connection
  to the recent resolution of the Poincar\'e conjecture, combinatorial and
  algebraic structures defined on surfaces are often relevant, e.g., via
  the study of mapping class groups and Teichm\"uller
  spaces~\cite{fm-pmcg-11};
\item in \emph{topological graph theory}, a branch of structural graph
  theory, graphs on surfaces are studied from a combinatorial point of
  view, also in relation to the theory of Robertson and Seymour on graph
  minors; for example, colorability questions of graphs on surfaces,
  generalizing the four-color theorem for planar graphs, are
  well-studied~\cite{mt-gs-01};
\item in \emph{enumerative combinatorics}, a natural problem is to count
  (exactly or asymptotically) maps with given properties in the plane or on
  surfaces, with the help of generating series; moreover, typical
  properties of random maps are
  investigated~\cite{m-trmag-09,b-tslpg-12,lz-gsta-04};
\item various \emph{applications} involve surface meshes, in particular in
  geometry processing and computer graphics, for
  approximation~\cite{cdp-hsmit-04}, topological
  simplification~\cite{gw-tnr-01,whds-retfi-04},
  compression~\cite{ag-rac3m-05}, and parameterization~\cite{gy-gcsp-03}.
  Techniques for general surfaces apply also to subsets of the plane, and
  are thus relevant in VLSI design~\cite{lm-artrp-85} and map
  simplification~\cite{bks-tcssu-98}.
\end{itemize}

This chapter is organized as follows.  We first review the basic concepts
and properties of topological surfaces and graphs embedded on them
(Sections~\ref{sec:surfaces} and~\ref{sec:graphs-surfaces}).
Then we consider three categories of topological
problems, mostly from a computational perspective: drawing an abstract
input graph on a surface (Section~\ref{sec:embedding}), homotopy questions and variations
(Section~\ref{sec:homotopy}), and optimization of curves and graphs on surfaces, also
from a homological point of view (Section~\ref{sec:optimalization}). Then we survey techniques
that allow us to solve general graph problems faster in the case where the
input graph is embedded on a fixed surface (Section~\ref{S:algo}).  Finally, we
collect other miscellaneous results (Section~\ref{sec:other-models}).

\A{SURFACES}\label{sec:surfaces}
\noindent
Surfaces are considered from a topological point of view: Two homeomorphic
surfaces are regarded as equivalent.  Surfaces such as the sphere or the
disk are topologically uninteresting; our focus is on surfaces in which
some closed curves are non-contractible (they cannot be deformed to a point
by a continuous motion on the surface).

\begin{figure}\psfrag{a}{$a$}\psfrag{b}{$b$}\psfrag{c}{$c$}\psfrag{d}{$d$}
  \centering
  \setlength{\tabcolsep}{3mm}
  {\footnotesize
  \begin{tabular}{p{.2\linewidth}|p{.2\linewidth}|p{.2\linewidth}|p{.2\linewidth}}
    disk (orientable, $g=0$, $b=1$) &
    sphere~$\SSS^2$ (orientable, $g=0$, $b=0$) &
    M\"obius strip (non-orientable, $g=1$, $b=1$) &
    handle (orientable, $g=1$, $b=1$)
    \\[2mm]
    \vcent{\includegraphics[width=\linewidth]{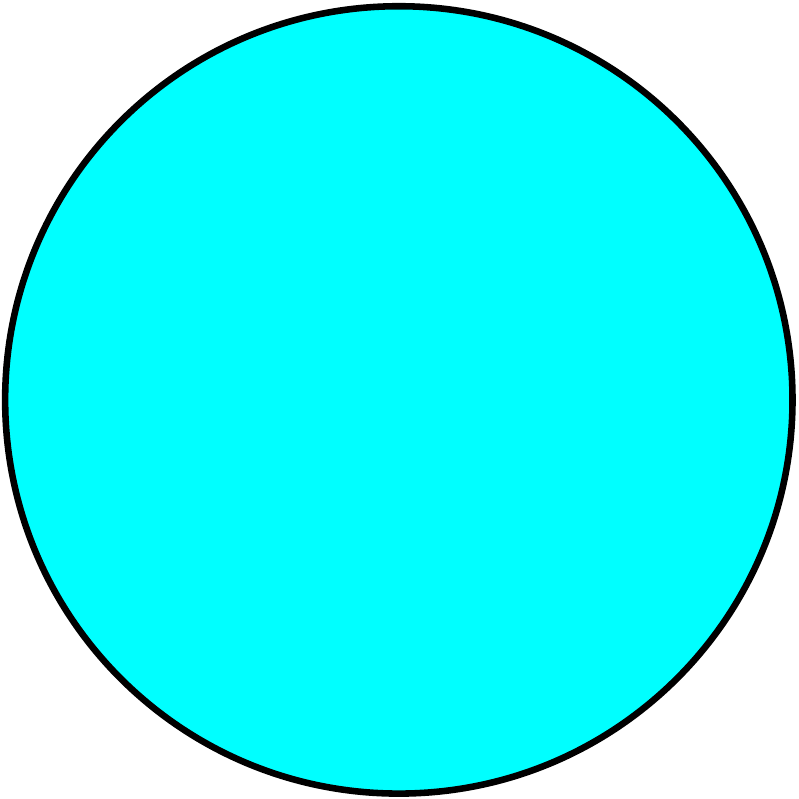}} &
    \vcent{\includegraphics[width=\linewidth]{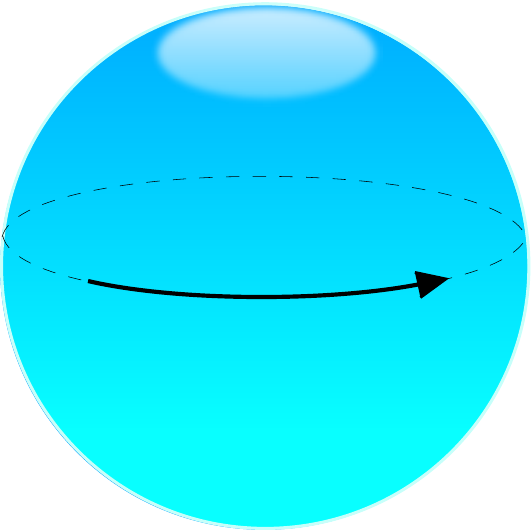}}%
    \ghost{\raisebox{.2ex}{\hspace{-4em}$a$}} &
    \vcent{\includegraphics[width=\linewidth,height=2cm]{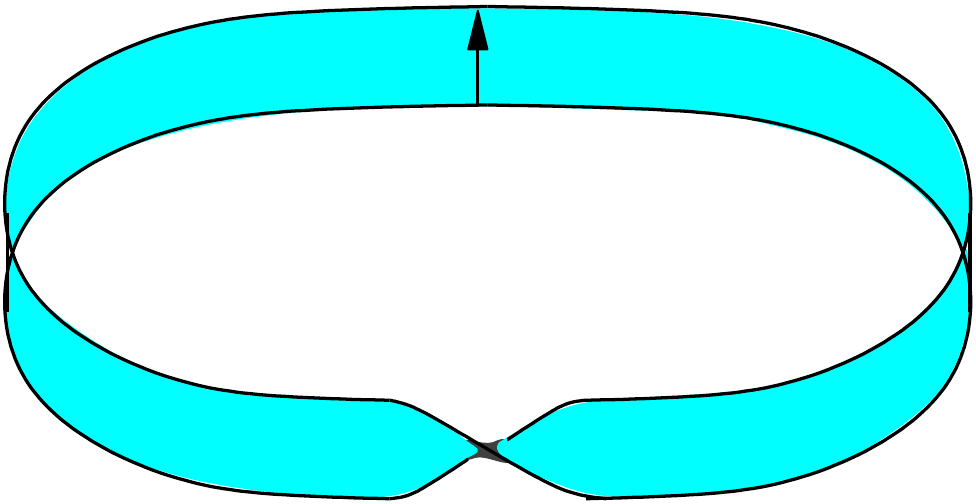}} &
    \vcent{\includegraphics[width=\linewidth]{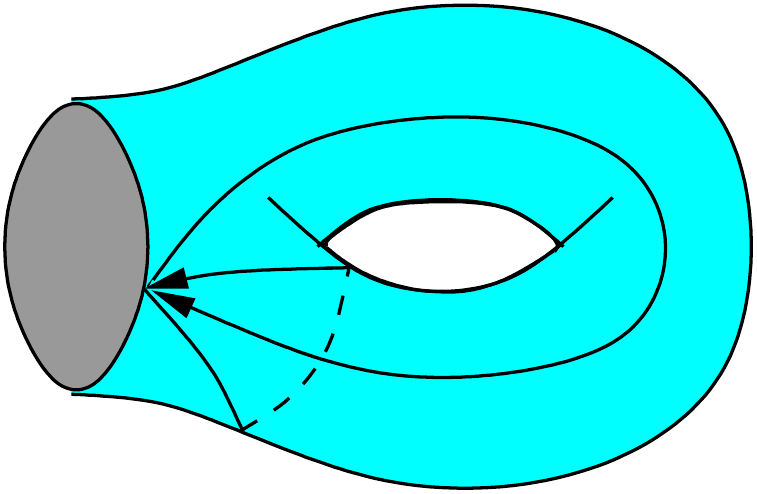}}
    \\[5mm]
    \vcent{\includegraphics[width=.9\linewidth]{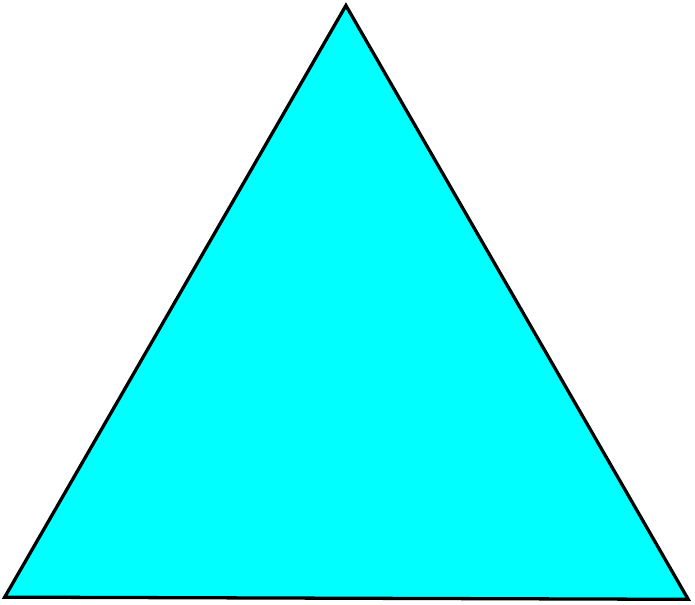}} &
    \centerline{\vcent{\includegraphics[width=.9\linewidth]{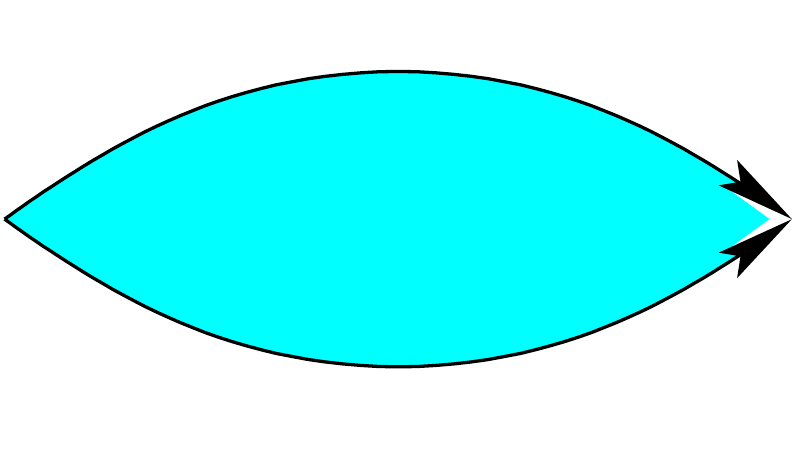}}} &
    \centerline{\vcent{\includegraphics[width=.9\linewidth]{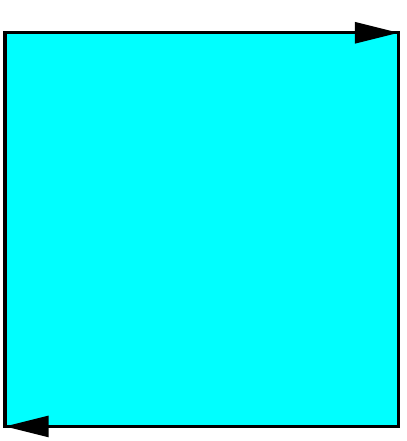}}} &
    \vcent{\includegraphics[width=.9\linewidth]{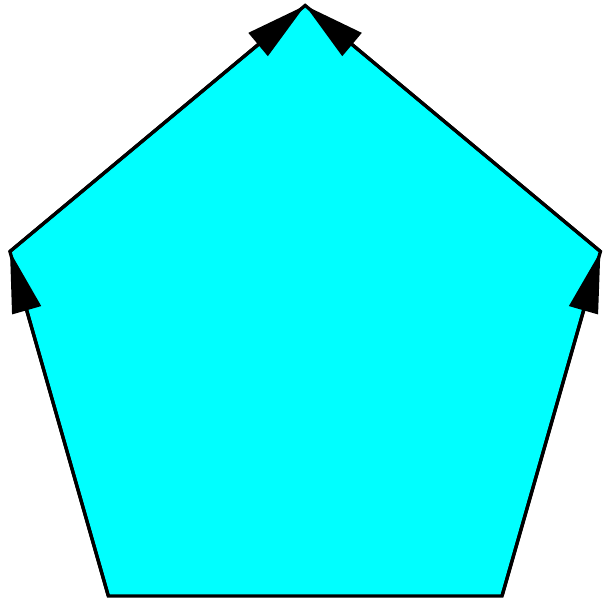}}\\
  \end{tabular}
}
\caption{Examples of surfaces.  Each surface (top row) comes with a
  polygonal schema (bottom row), a polygon with some labeled and directed
  edges; the surface can be obtained by identifying the pairs of edges with
  the same labels, respecting their direction.  The genus~$g$ and number of
  boundary components~$b$ are specified, as well as whether the surface is
  orientable.}
  \label{F:surfaces}
\end{figure}

\Bnn{GLOSSARY}
\begin{gllist}
\item {\trmbitx Homeomorphism:}\tindex{homeomorphism}\quad
  Given two topological spaces $X$
  and~$X'$, a map $h\colon X\to X'$ is a homeomorphism if $h$ is bijective
  and both $h$ and its inverse are continuous.
\item {\trmbitx Surface (topological definition):}\tindex{surface}\quad
  In this chapter, a
  surface~$\surf$ is a \emph{compact} two-dimensional manifold possibly
  with boundary.  Equivalently, $\surf$ is a compact topological space that is
  Hausdorff (any two distinct points have disjoint neighborhoods) and such
  that every point has a neighborhood homeomorphic to the plane or the
  closed half-plane.  The set of points of a surface~$\surf$ that have no
  neighborhood homeomorphic to the plane is the {\trmbitx boundary}\tindex{boundary!of surface}
  of~$\surf$.
\item {\trmbitx Surface (combinatorial definition):}\quad Equivalently, a
  surface~$\surf$ is a topological space obtained from finitely many
  disjoint triangles by identifying some pairs of edges of the triangles
  (by the quotient topology).  The {\trmbitx boundary} of~$\surf$ is the
  union of the edges that are not identified with any other edge.
\item {\trmbitx Path:}\quad A path on~$\surf$ is a continuous map $p\colon
  [0,1]\to\surf$.  Its two {\trmbitx endpoints} are $p(0)$ and $p(1)$.
\item {\trmbitx Connectedness:}\tindex{connectedness}\quad
  A surface is {\trmbitx connected} if
  any two points of the surface are the endpoints of some path.  The
  inclusionwise maximal connected subsets of a surface form its {\trmbitx
    connected components}.
\item {\trmbitx Orientability:}\tindex{orientable!surface}\quad
  A surface is {\trmbitx non-orientable}
  if some subset of it (with the induced topology) is homeomorphic to the
  M\"obius strip (defined in Figure~\ref{F:surfaces}).  Otherwise, it is
  {\trmbitx orientable}.
\end{gllist}

\Bnn{PROPERTIES: CLASSIFICATION OF SURFACES}
\noindent
Every connected surface is homeomorphic to exactly one of the following
surfaces:
\begin{itemize}
\item the orientable surface of {\trmbitx genus}~$g\ge0$ with $b\ge0$
  {\trmbitx boundary components}\tindex{boundary!component} (or, more concisely, {\trmbitx
    boundaries}), obtained from the sphere by removing $g$~disjoint open
  disks, attaching a handle (defined in Figure~\ref{F:surfaces}) to each of
  the resulting $g$~circles, and finally removing $b$~open disks with
  disjoint closures;
\item the non-orientable surface of {\trmbitx genus}~$g\ge1$ with $b\ge0$
  {\trmbitx boundary components} (or {\trmbitx boundaries}), obtained from
  the sphere by removing $g$~disjoint open disks, attaching a M\"obius
  strip (defined in Figure~\ref{F:surfaces}) to each of the resulting
  $g$~circles, and finally removing $b$~open disks with disjoint closures.
\end{itemize}

\begin{figure}[h]
  \centering
  \includegraphics[width=\linewidth]{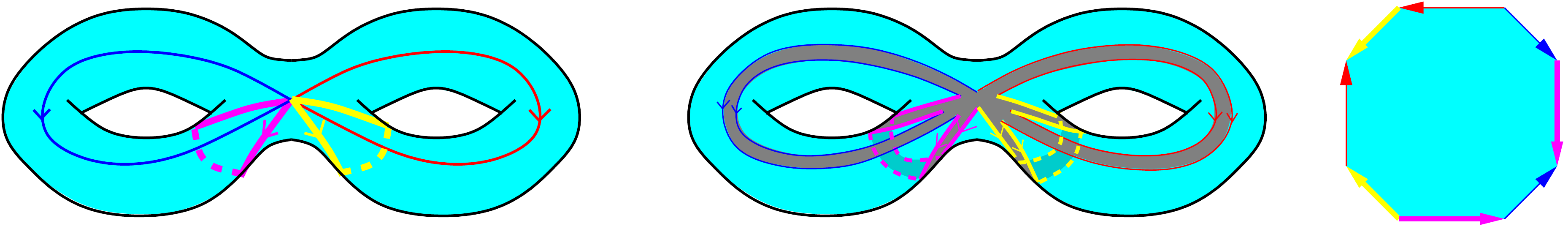}
  \caption{A double torus with a system of loops (left); the surface cut
    along the loops (middle) is a disk, shown in the form of a (canonical)
    polygonal schema (right).}
  \label{F:schema}
\end{figure}

Every surface can be obtained by identifying pairs of edges of disjoint
triangles.  More concisely, every surface can be defined by a {\trmbitx
  polygonal schema}\tindex{polygonal schema}\tindex{schema!polygonal},
  a polygon with labels and directions on some of the
edges specifying how they must be identified.  In particular, one can
define a {\trmbitx canonical}\tindex{canonical polygonal schema} polygonal schema
for every connected surface without boundary:
\begin{itemize}
\item The canonical polygonal schema of the orientable surface of genus
  $g\ge 1$ is a $4g$-gon whose successive edges are labeled
  $a_1,b_1,\bar a_1,\bar b_1, \ldots, a_g,b_g,\bar a_g,\bar b_g$, and where
  edge $x$ is directed clockwise, edge $\bar x$ is directed
  counterclockwise.  Identifying edge $x$ with $\bar x$, as indicated by
  their directions, gives the orientable surface of genus~$g$.  See
  Figure~\ref{F:schema}.
\item Similarly, the canonical polygonal schema of the non-orientable
  surface of genus $g\ge 1$ is a $2g$-gon whose successive edges are
  labeled $a_1,a_1,\ldots, a_g,a_g$, and where all edges are directed
  clockwise.
\end{itemize}

\Bnn{EXAMPLES}

\noindent
Table~\ref{T:surfaces} lists some common connected surfaces.  See also Figures \ref{F:surfaces}
and~\ref{F:schema}.

\begin{table}
  \renewcommand{\arraystretch}{.9}
  {%
  \begin{tabular}{c|ccc}
    Surface & Orientable? & Genus & \# of boundary components\\\hline
    Sphere & Yes & 0 & 0\\
    Disk & Yes & 0 & 1\\
    Annulus = cylinder & Yes & 0 & 2\\
    Pair of pants & Yes & 0 & 3\\\hline
    Torus & Yes & 1 & 0\\
    Handle & Yes & 1 & 1\\\hline
    Double torus & Yes & 2 & 0\\\hline\hline
    Projective plane & No & 1 & 0\\
    M\"obius strip & No & 1 & 1\\\hline
    Klein bottle & No & 2 & 0\\
  \end{tabular}}
  \caption{Some common surfaces.}
  \label{T:surfaces}%
\end{table}

\A{GRAPHS ON SURFACES}\label{sec:graphs-surfaces}

\Bnn{GLOSSARY}

\noindent Let $\surf$ be a surface.

\smallskip

\begin{gllist}
\item {\trmbitx Loop:}\tindex{loop}\quad A loop is a path whose two endpoints are equal
  to a single point, called the {\trmbitx basepoint}\tindex{basepoint} of the loop.
\item {\trmbitx Closed curve:}\tindex{closed curve}\tindex{curve!closed}\quad
   A closed curve on~$\surf$ is a
  continuous map from the unit circle~$\SSS^1$ to~$\surf$.  This is almost
  the same as a loop, except that a closed curve has no distinguished
  basepoint.  A closed curve is sometimes called a {\trmbitx cycle},
  although, contrary to the standard terminology in graph theory, here a
  cycle may self-intersect.
\item {\trmbitx Curve:}\tindex{curve}\quad A curve is either a path or a closed curve.
  For most purposes, the parameterization is unimportant; for example, a
  path~$p$ could be regarded as equivalent to $p\circ\varphi$, where
  $\varphi:[0,1]\to [0,1]$ is bijective and increasing.
\item {\trmbitx Simplicity:}\tindex{curve!simple}\quad
   A path or a closed curve is {\trmbitx
    simple} if it is injective.  A loop $\ell\colon[0,1]\to\surf$ is
  {\trmbitx simple} if its restriction to $[0
  ,1)$ is injective.
\item {\trmbitx Graph:}\tindex{graph}\quad In this chapter, unless specified otherwise,
  graphs are finite, undirected, and may have loops and multiple edges.
\item {\trmbitx Curve (in a graph):}\quad A {\trmbitx curve} in a graph~$G$
  (also called {\trmbitx walk}\tindex{walk} in the terminology of graph theory) is a
  sequence of directed edges $e_1,\ldots, e_k$ of~$G$ where the target
  of~$e_i$ equals the source of~$e_{i+1}$.  Repetitions of vertices and
  edges are allowed.  The {\trmbitx endpoints} of the curve are the source
  of~$e_1$ and the target of~$e_k$.  If they are equal, the curve is
  {\trmbitx closed}.
\item {\trmbitx Graph embedding (topological definition):} \tindex{graph
    embedding}\tindex{embedding!of a graph}\quad A graph~$G$ naturally
  leads to a topological space~$\hat G$, defined as follows: One considers
  a disjoint set of segments, one per edge of~$G$, and identifies the
  endpoints that correspond to the same vertex of~$G$.  This gives a
  topological space, from which $\hat G$ is obtained by adding one isolated
  point per isolated vertex of~$G$.  (As a special case, if $G$ has no loop
  and no multiple edge, then $G$ is a one-dimensional simplicial complex,
  and $\hat G$ is the associated topological space.)  An {\trmbitx
    embedding} of~$G$ is a continuous map from~$\hat G$ into~$\surf$ that
  is a homeomorphism from~$\hat G$ onto its image.
\item {\trmbitx Graph embedding (concrete definition):}\quad Equivalently,
  an embedding of~$G$ on~$\surf$ is a ``crossing-free'' drawing of~$G$: It
  maps the vertices of~$G$ to distinct points of~$\surf$, and its edges to
  paths of~$\surf$ whose endpoints are the images of their incident
  vertices; the image of an edge can self-intersect, or intersect the image
  of another edge or vertex, only at its endpoints.  When no confusion
  arises, we identify~$G$ with its embedding on~$\surf$, or with the image
  of that embedding.
\item {\trmbitx Face:}\tindex{face!of embedded graph}\quad
    The faces of an embedded graph~$G$ are the
  connected components of the complement of the image of~$G$.
\item {\trmbitx Degree:}\tindex{degree!of a vertex}\quad The degree of a vertex~$v$ is the number of
  edges incident to~$v$, counted with multiplicity (if an edge is a loop).
  The degree of a face~$f$ is the number of edges incident to~$f$, counted
  with multiplicity (if an edge has the same face on both sides).
\item {\trmbitx Cellular embedding:}
   \tindex{cellular embedding}\tindex{embedding!cellular}\quad
   A graph embedding is {\trmbitx cellular} if its faces are homeomorphic to open disks.
\item {\trmbitx Triangulation:}\tindex{triangulation!of embedded graphs}\quad
  A graph embedding is a triangulation
  if it is cellular and its faces have degree three.  The triangulation may
  fail to be a simplicial complex: A triangle is not necessarily incident
  to three distinct vertices, or even to three distinct edges.
\item {\trmbitx Cutting:}\tindex{cutting!of embedded graphs}\quad
  Given an embedded graph~$G$ on~$\surf$
  without isolated vertex, the operation of cutting $\surf$ along~$G$
  results in a (possibly disconnected) surface with boundary, denoted
  $\surf\cut G$ (or sometimes $\surf$\Rightscissors$G$ or similar); each
  connected component of $\surf\cut G$ corresponds to a face of~$G$
  on~$\surf$, and by identifying pieces of the boundaries of these
  components in the obvious way, one recovers the surface~$\surf$.
  Similarly, one can cut along a set of disjoint, simple closed curves.
  (Technically, if $\surf$ has non-empty boundary, an additional condition
  is needed: The intersection of an edge with the boundary of~$\surf$ can
  be either the entire edge, its two endpoints, or one of its endpoints.)
\item {\trmbitx Planarity:}\index{planar graph}\tindex{graph!planar}\quad
  A graph is planar if it has an embedding
  to the plane (or equivalently the sphere).
\item {\trmbitx Dual graph:}\tindex{dual graph}\tindex{graph!dual}\quad
  A dual graph of a cellularly embedded
  graph~$G$ on~$\surf$ (assumed without boundary) is a graph~$G^*$ embedded
  on~$\surf$ with one vertex~$f^*$ inside each face~$f$ of~$\surf$, and
  with an edge~$e^*$ for each edge~$e$ of~$G$, such that $e^*$ crosses $e$
  and no other edge of~$G$.  A dual graph is cellularly embedded.  Its
  combinatorial map (see below) is uniquely determined by the combinatorial
  map of~$G$.
\item {\trmbitx Euler genus:}\tindex{Euler genus}\tindex{genus!Euler}\quad
  The Euler genus~$\bar g$ of a connected
  surface~$\surf$ with genus~$g$ equals~$2g$ if $\surf$~is orientable,
  and~$g$ if $\surf$~is non-orientable.
\item {\trmbitx Euler characteristic:}
  \tindex{Euler characteristic}\tindex{characteristic!Euler}\quad
  The Euler characteristic of a
  cellularly embedded graph~$G$ equals $\chi(G):=v-e+f$, where $v$, $e$,
  and~$f$ are its number of vertices, edges, and faces, respectively.
\end{gllist}

\Bnn{PROPERTIES: EULER'S FORMULA AND CONSEQUENCES}

\begin{enumerate}
\item{\trmbitx Euler's formula:} If $G$ is \emph{cellularly} embedded on a
  connected surface~$\surf$ of Euler genus~$\bar g$ with $b$~boundary
  components, then $\chi(G)=2-\bar{g}-b$.  In particular, $\chi(G)$ does
  not depend on~$G$, only on~$\surf$, and is consequently called the
  {\trmbitx Euler characteristic} of~$\surf$.
\item The number of vertices and faces of a graph~$G$ cellularly embedded
  on a connected surface is at most linear in the number of its edges.  In
  particular, the combinatorial complexity of~$G$ is linear in its number
  of edges.
\item Conversely, let $G$ be a (not necessarily cellular) graph embedding
  on a connected surface with Euler genus~$\bar g$ and $b$ boundaries.
  Assume that $G$~has no face of degree one or two that is an open disk.
  Then the numbers $e$ of edges and $v$ of vertices of~$G$ satisfy
  $e=O(v+\bar g+b)$.
\end{enumerate}

\Bnn{DATA STRUCTURES}
\noindent
In all the problems we shall consider, the exact embedding of a graph on a
surface is irrelevant; only the actual combinatorial data associated to the
embedding is meaningful.  If $G$ is a graph cellularly embedded on a
surface~$\surf$ without boundary, we only need the information of~$G$
together with the \emph{facial walks}\tindex{walk!facial}\tindex{facial walk},
namely, the closed walks in~$G$
encountered when walking along the boundary of the faces of~$G$.  This
information is called the
\emph{combinatorial map}\tindex{combinatorial!map}\tindex{map!combinatorial}
of~$G$, and allows us to reconstruct the surface, by attaching disks to every
facial walk. (Some conditions on the walks are needed to ensure that the resulting
space is indeed a surface.)  If $\surf$ has boundaries, one can specify the
corresponding faces of~$G$.  If $\surf$ is orientable and $G$ has no loop
edge, instead of the facial walks one could as well specify the cyclic
ordering of the edges incident to each vertex.

However, more complicated data structures are needed to perform basic
operations efficiently.  For example, one should be able to compute the
degree of a face in time linear in the degree; to count the number of faces
of~$G$ in linear time; to determine whether the surface is orientable in
linear time; etc.  (The last two operations, together with counting the
number of vertices and edges, allow us to identify the topology of the surface
in linear time using Euler's formula; this can also be done in logarithmic
space~\cite{bekt-2mrl-16}.)

\begin{figure}[h]
  \centering
  \includegraphics[width=.48\linewidth]{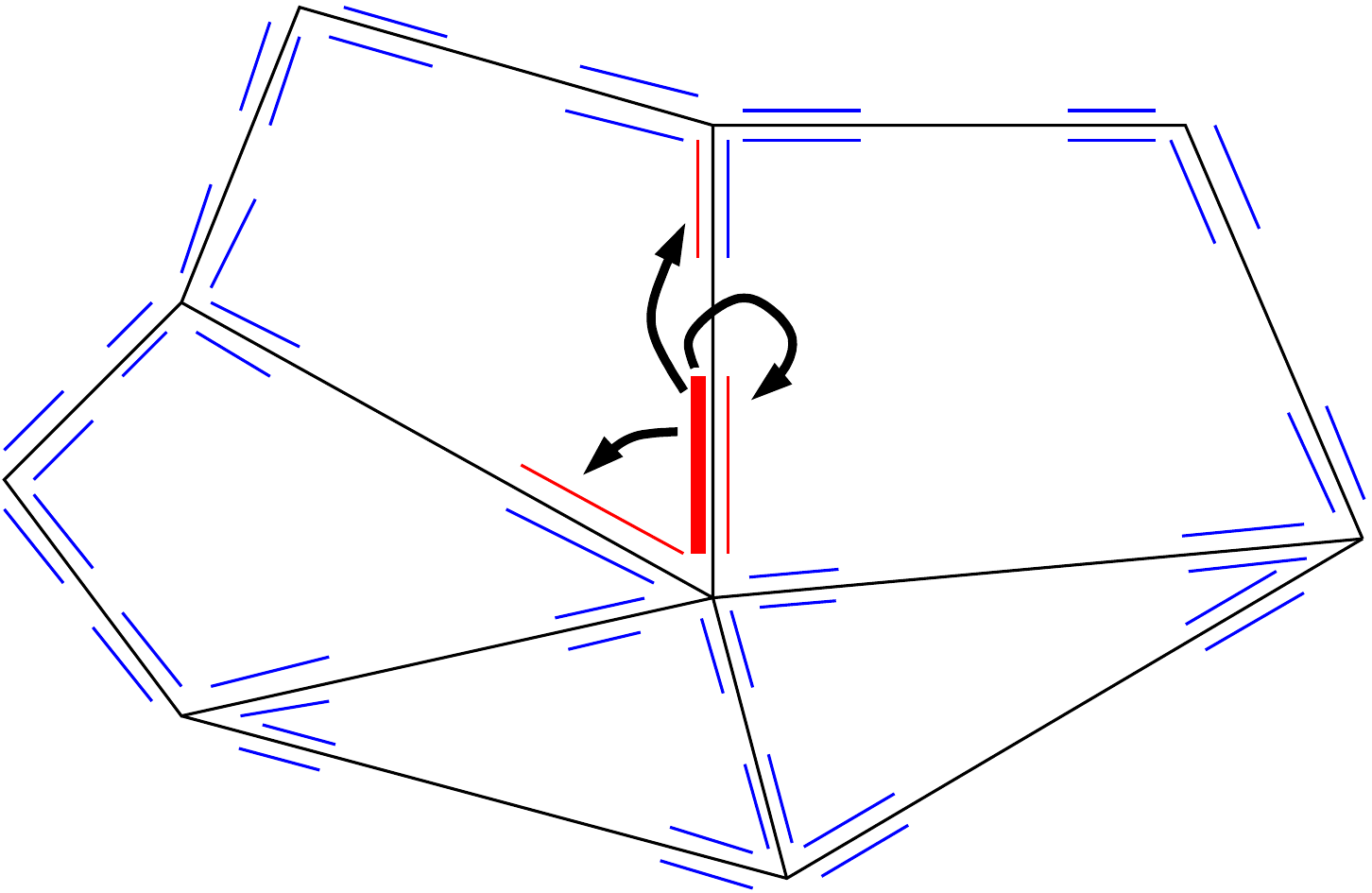}
  \caption{The graph-encoded map data structure.  Each edge bears four
    flags, drawn parallel to it.  Three operations allow us to move from a
    flag to a nearby flag.}
  \label{F:flags}
\end{figure}

One such data structure, the \emph{graph-encoded map}\tindex{graph-encoded map}\tindex{map!graph-encoded}
or \emph{gem representation}\tindex{gem representation}~\cite{l-gem-82}, uses flags\tindex{flag}
(quarter-edges, or, equivalently, incidences between a vertex, an edge, and a face) of~$G$, see
Figure~\ref{F:flags}; three involutive operations can be applied to a flag to move to
an incident flag.  Alternative data structures have been designed for more
general situations (e.g., to allow surfaces with boundaries) or to take
advantage of special situations (e.g., in the case where $G$ is a
triangulation, or where $\surf$ is orientable); see the
survey~\cite{k-ugpdd-99}.  However, the choice of the data structure is
irrelevant for the theoretical design and asymptotic analysis of the
algorithms.

\Bnn{CONVENTIONS FOR THIS CHAPTER}
\noindent
Henceforth, \textbf{we assume all surfaces to be connected}.

In several works mentioned in the following, only orientable surfaces
are considered.  In some cases, non-orientable surfaces are just as easy to
handle, but sometimes they lead to additional difficulties. 
\textbf{We refer to the original articles to determine whether the results 
hold  
on non-orientable surfaces.}

Also, in most problems studied in this chapter, surfaces with boundaries are no
harder to handle than surfaces without boundary: Any algorithm for surfaces
without boundary immediately implies an algorithm for surfaces with
boundary (with the same running time, or by replacing $g$ by $g+b$ in the
complexity, where $g$ and~$b$ are the genus and the number of boundary
components).  For this reason, \textbf{we mostly focus on computational
  problems for surfaces without boundary}.

Finally, when we consider cellularly embedded graphs for algorithmic
problems, \textbf{we implicitly assume that they are specified in the form
  of a data structure as described above} (e.g., a graph-encoded map).

\A{EMBEDDING AND DRAWING GRAPHS ON SURFACES}\label{sec:embedding}
\noindent
Being able to build embeddings of a graph on a surface with small genus is
important; almost all algorithms for graphs embeddable on a fixed surface
require an embedding of the input graph (there are a few
exceptions~\cite{es-noaaa-14,k-spebc-06,ms-pus-12}).  We discuss
algorithmic results related to the problem of embedding a graph on a
surface, and then consider more general drawings where crossings are
allowed.

\Bnn{EMBEDDING GRAPHS ON SURFACES}
\noindent
Let $G$ be an abstract graph (not embedded on any surface), given, e.g., by
the (unordered) list of the edges incident to every vertex.  We assume that
$G$ is connected.  Let $n$ denote the combinatorial complexity of~$G$, that
is, the total number of vertices and edges of~$G$.
\begin{enumerate}
\item\emph{General facts:} An embedding on an orientable surface with
  minimum possible genus is cellular.  If $G$ is embeddable on an
  orientable (resp., non-orientable) surface of genus~$g$, then it is
  embeddable on an orientable (resp., non-orientable) surface of
  genus~$g'$, for every $g'\ge g$.
\item\emph{General bound:} $G$ can be cellularly embedded on some
  orientable surface with genus~$O(n)$.
\item\emph{Planar case:} There is an $O(n)$-time algorithm for deciding
  embeddability in the sphere (equivalently, in the
  plane)~\cite{ht-ept-74}; also in $O(n)$ time, the graph can be embedded
  with straight-line segments in the plane~\cite{s-epgg-90} (see
  also~\cite[Ch.~4]{nr-pgd-04}), if it has no loop or multiple edge.
  See Chapter~\chapGraphDraw\ for more results on graph drawing.
\item\emph{Time complexity:} Given a graph~$G$ and a surface~$\surf$,
  specified by its Euler genus~$\bar g$ and by whether it is orientable,
  determining whether $G$ embeds on~$\surf$ is NP-hard~\cite{t-ggpnc-89},
  but can be done in $2^{\poly(\bar g)}\cdot n$
  time~\cite{kmr-sltae-08,m-ltaeg-99} (where $\poly(\bar g)$ is a
  polynomial in~$\bar g$), which is linear if $\bar g$ is fixed.  Such an
  embedding can be computed in the same amount of time if it exists.
\item\emph{Space complexity:} For every \emph{fixed} $\bar g$, determining
  whether an input graph~$G$ embeds on some surface (orientable or not) of
  Euler genus at most~$\bar g$ can be done in space logarithmic in the
  input size~\cite{ek-ecgbg-14}.
\item\emph{Approximation:} Given as input a graph~$G$ and an integer~$\bar
  g$, one can in polynomial time either correctly report that $G$ embeds on
  no surface of Euler genus~$\bar g$, or compute an embedding on some
  surface of Euler genus $\bar g^{O(1)}$~\cite{ks-becag-15}.
\end{enumerate}

Except for the planar case, these algorithms are rather complicated, and
implementing them is a real challenge.  For example, there seems to be no
available implementation of a polynomial-time algorithm for testing
embeddability in the torus, and no publicly available implementation of any
algorithm to decide whether a graph embeds on the double torus; attempts of
implementing some known embedding algorithms, even in the simplest cases,
have unveiled some difficulties~\cite{mk-egea-11}.  On the other hand, a
recent approach is promising in practice for graphs of moderate size, using
integer linear programming or Boolean satisfiability
reformulations~\cite{bchk-pmmgg-16}.

In contrast, determining the \emph{maximum} genus of an orientable surface
without boundary on which a graph can be \emph{cellularly} embedded can be
done in polynomial time~\cite{fgm-fmggi-88}.  There are also results on the
embeddability of two-dimensional simplicial complexes on
surfaces~\cite{m-mg2c-97}.

On a less algorithmic side, in the field of topological graph theory, a lot
more is known about the embeddability of some classes of graphs on some
surfaces; see, e.g.,~\cite[Sect.~4.2]{a-tgts-96} and references therein.

\Bnn{GLOSSARY ON DRAWINGS}

\noindent
Let $G$ be a graph and $\surf$ be a surface.

\smallskip

\begin{gllist}
\item {\trmbitx Drawing:}\tindex{drawing!graph}\quad
   Drawings are more general than embeddings in
  that they allow a finite set of crossing points, where exactly two pieces
  of edges intersect and actually cross.  Formally, recall that $G$ has an
  associated topological space~$\hat G$.  A (topological) {\trmbitx
    drawing} of~$G$ on~$\surf$ is a continuous map from $\hat G$
  into~$\surf$ such that the preimage of every point in~$\surf$ has
  cardinality zero or one, except for a finite set of points
  (``crossings''), whose preimages have cardinality two; moreover, each
  such crossing point has a disk neighborhood that contains exactly the
  images of two pieces of edges of~$\hat G$, which form, up to
  homeomorphism, two crossing straight lines.
\item {\trmbitx Arrangement:}\tindex{arrangement}\quad
  Let $D$ be a drawing of~$G$ on~$\surf$.
  The {\trmbitx arrangement} of~$D$ on~$\surf$ is the graph~$G'$ embedded
  on~$\surf$ that has the same image as~$D$ and is obtained from~$D$ by
  inserting a vertex of degree four at each crossing in~$D$ and subdividing
  the edges of~$G$ accordingly.  Similarly, one can consider the
  arrangement of a set of curves drawn on~$\surf$.
\item {\trmbitx Crossing number:}\tindex{crossing number}\tindex{graph!crossing number of}\quad
  The crossing number of~$G$ with
  respect to~$\surf$ is the minimum number of crossings that $G$ has in any
  drawing of~$G$ on~$\surf$.
\item {\trmbitx Pair crossing number:}\tindex{pair crossing number}\tindex{graph!pair crossing number of}
  \quad The pair crossing number of~$G$
  with respect to~$\surf$ is the minimum number of pairs of edges of~$G$
  that cross, over all drawings of~$G$ on~$\surf$.
\item {\trmbitx Odd crossing number:}\tindex{odd crossing number}\tindex{graph!odd crossing number of}
  \quad The odd crossing number of~$G$
  with respect to~$\surf$ is the minimum number of pairs of edges of~$G$
  that cross an odd number of times, over all drawings of~$G$ on~$\surf$.
\end{gllist}

\Bnn{DRAWING GRAPHS ON SURFACES WITH FEW CROSSINGS}

\begin{enumerate}
\item \emph{Crossing numbers:} Computing the planar crossing number of a
  graph is NP-hard (even in very special cases, such as that of a planar
  graph with a single additional edge~\cite{cm-aoepg-13}), and there exists
  no polynomial-time algorithm with approximation guarantee better than a
  certain constant~\cite{c-hacn-13}.  However, for every fixed~$k$, one
  can, in linear time, determine whether an input graph has planar crossing
  number at most~$k$~\cite{kr-ccnlt-07}, although the problem admits no
  polynomial kernel~\cite{hd-cnhk-16}.  Some approximation algorithms for
  the planar crossing number are known in restricted cases, such as bounded
  maximum degree~\cite{hc-acnge-10,c-agcnp-11}.
\item \emph{Variations on crossing numbers:} The relations between the
  various notions of crossing numbers are not fully understood.  Let $c$,
  $p$, and~$o$ denote the planar crossing number, planar pair crossing
  number, and planar odd crossing number, respectively, of some graph~$G$.
  It is clear that $o\le p\le c$, and it is known that the left inequality
  can be strict~\cite{pss-ocncn-08}.  It is widely believed that $p=c$, but
  the best bound known so far is $c=O(p^{3/2}\log^2p)$ (this follows
  essentially from~\cite{t-bbpcn-12}).  See, e.g.,~\cite{m-sgs-13} for more
  details, and~\cite{s-gcnvs-13} for a wide survey on the various notions
  of crossing numbers.
\item \emph{Hanani--Tutte theorem:} The (weak) Hanani--Tutte
  theorem~\cite{h-uwukd-34,t-ttcn-70}, however, states that if $o=0$
  then~$c=0$.  Furthermore it holds not only for the plane, but for
  arbitrary surfaces~\cite{cn-bgt-00,pss-recs-09}: If a graph~$G$ can be
  drawn on a surface~$\surf$ in a way that every pair of edges crosses an
  even number of times, then $G$ can be embedded on~$\surf$.  In the planar
  case, it actually suffices to assume that every pair of
  \emph{independent} edges (which do not share any endpoints) crosses an
  even number of times, but whether this generalizes to arbitrary surfaces
  is open, except for the projective
  plane~\cite{pss-shtpp-09,ckppt-dpsht-16}.  We refer to
  surveys~\cite{s-ttpht-13,s-htrr-14} for more details.
\end{enumerate}

\A{HOMOTOPY AND ISOTOPY}\label{sec:homotopy}
\noindent
Most works in computational topology for surfaces do not take as input a
given abstract graph, as in the previous section; instead, they consider an
already embedded graph, given by its combinatorial map.

\Bnn{GLOSSARY}

\noindent Let $\surf$ be a surface.

\smallskip

\begin{gllist}
\item{\trmbitx Reversal:}\tindex{reversal of a path}\tindex{reverse path}\quad
  The reversal of a path $p\colon
  [0,1]\to\surf$ is the path $p\inv\colon[0,1]\to\surf$ defined by
  $p\inv(t)=p(1-t)$.
\item{\trmbitx Concatenation:}\tindex{concatenation!of paths}\quad
  The concatenation of two paths
  $p,q\colon[0,1]\to\surf$ with $p(1)=q(0)$ is the path $p\cdot q$ defined
  by $(p\cdot q)(t)=p(2t)$ if $t\le1/2$ and $(p\cdot q)(t)=q(2t-1)$ if
  $t\ge1/2$.
\item{\trmbitx Homotopy for paths:}\tindex{homotopy!of paths}\quad Given two paths $p,q\colon
  [0,1]\to\surf$, a {\trmbitx homotopy} between $p$ and~$q$ is a continuous
  deformation between $p$ and~$q$ that keeps the endpoints fixed.  More
  formally, it is a continuous map $h\colon [0,1]\times [0,1]\to\surf$ such
  that $h(0,\cdot)=p$, $h(1,\cdot)=q$, and both $h(\cdot,0)$ and
  $h(\cdot,1)$ are constant maps (equal, respectively, to $p(0)=q(0)$ and
  to $p(1)=q(1)$).  The paths $p$ and~$q$ are {\trmbitx homotopic}.  Being
  homotopic is an equivalence relation, partitioning the paths with given
  endpoints into {\trmbitx homotopy classes}\tindex{homotopy class}.
\item{\trmbitx Fundamental group:}\tindex{fundamental group}\tindex{group!fundamental}\quad
  The homotopy classes of loops with
  a given basepoint form a group, where concatenation of loops accounts for
  the multiplication and reversal accounts for the inverse operation: if
  $[p]$ denotes the homotopy class of path~$p$, then we have
  $[p\cdot q]=[p]\cdot [q]$ and $[p\inv]=[p]\inv$.
\item{\trmbitx Homotopy for closed curves} (also called {\trmbitx free
    homotopy}):\tindex{free homotopy}\quad Given two closed curves
  $\gamma,\delta\colon\SSS^1\to\surf$, a {\trmbitx homotopy} between $\gamma$
  and~$\delta$ is a continuous deformation between them, namely, a
  continuous map $h\colon [0,1]\times\SSS^1\to\surf$ such that
  $h(0,\cdot)=\gamma$ and $h(1,\cdot)=\delta$.
\item{\trmbitx Contractibility:}\tindex{contractible}\quad A loop or closed curve is {\trmbitx
    contractible} if it is homotopic to a constant loop or closed curve.
\item{\trmbitx Isotopy:}\tindex{isotopy!of a path}\tindex{isotopy!of a graph}\quad
  An isotopy between two \emph{simple} paths,
  loops, or closed curves is a homotopy~$h$ that does not create
  self-intersections: for each~$t$, $h(t,\cdot)$ is a simple path, loop, or
  closed curve.  An isotopy of a graph~$G$ is a continuous family of
  embeddings of~$G$ (the vertices and edges move continuously).
\item{\trmbitx Ambient isotopy:}\tindex{isotopy!ambient}\quad An ambient isotopy of a
  surface~$\surf$ is a continuous map $i:[0,1]\times\surf\to\surf$ such
  that for each $t\in[0,1]$, $i(t,\cdot)$ is a homeomorphism.
\item{\trmbitx Minimally crossing:}\tindex{minimally crossing!curves}\quad A family of closed
  curves~$\Gamma=(\gamma_1,\ldots,\gamma_k)$ is minimally crossing if for
  every family of closed curves $\Gamma'=(\gamma'_1,\ldots,\gamma'_k)$ with
  $\gamma_i$ and~$\gamma'_i$ homotopic for each~$i$, the number of
  intersections and self-intersections in~$\Gamma$ is no larger than
  in~$\Gamma'$.
\item{\trmbitx Covering space:}\tindex{covering space}\quad Let $\tilde\surf$ be a possibly
  non-compact connected surface.  A continuous map
  $\pi\colon\tilde\surf\to\surf$ is a {\trmbitx covering map} if every
  point $x\in\surf$ has a connected neighborhood~$U$ such that $\pi\inv(U)$
  is a disjoint union of open sets $(U_i)_{i\in I}$ and $\pi|_{U_i}\colon
  U_i\to U$ is a homeomorphism for each~$i$.  We say that
  $(\tilde\surf,\pi)$ is a {\trmbitx covering space} of~$\surf$.  A
  {\trmbitx lift}\tindex{lifting} of a path~$p$ is a path~$\tilde p$ on~$\tilde\surf$ such
  that $\pi\circ\tilde p=p$.  Finally, if each loop in~$\tilde\surf$ is
  contractible, then $(\tilde\surf,\pi)$ is a {\trmbitx universal covering
   space}\tindex{covering space!universal} of~$\surf$, which is essentially unique
  (precisely: if $(\tilde\surf,\pi)$ and $(\tilde\surf',\pi')$ are universal covering
  spaces, then there is a homeomorphism $\tau:\tilde\surf\to\tilde\surf'$
  such that $\pi=\pi'\circ\tau$).
\end{gllist}

\Bnn{BASIC PROPERTIES}

\begin{enumerate}
\item Two paths $p$ and~$q$ are homotopic if and only if $p\cdot q\inv$ is
  a (well-defined and) contractible loop.
\item Two loops $p$ and~$q$ with the same basepoint are freely homotopic
  (viewed as closed curves without basepoint) if the homotopy classes of
  the loops $p$ and~$q$ are conjugates in the fundamental group.
\item The fundamental group of a surface~$\surf$ without boundary of
  genus~$g$ is best understood by looking at a canonical polygonal schema
  of the surface: If $\surf$~is orientable, it is the group generated by
  $2g$~generators $a_1,b_1,\ldots,a_g,b_g$ and with a single relation,
  $a_1b_1a_1\inv b_1\inv\ldots a_gb_ga_g\inv b_g\inv$, corresponding to the
  boundary of the polygonal schema.  Similarly, if $\surf$ is
  non-orientable, it is the group generated by $g$~generators
  $a_1,\ldots,a_g$ and with a single relation, $a_1a_1\ldots a_ga_g$.
\item The fundamental group of a surface with at least one boundary
  component is a free group (because such a surface has the homotopy type
  of a graph).
\item Let $(\tilde\surf,\pi)$ be a covering space of~$\surf$.  Every
  path~$p$ on~$\surf$ admits lifts on~$\tilde\surf$; moreover, if $\tilde
  x$ is a lift of~$p(0)$, then $p$ has a unique lift~$\tilde p$ such that
  $\tilde p(0)=\tilde x$.  Two paths are homotopic on~$\surf$ if and only
  if they have homotopic lifts on~$\tilde\surf$.  In particular, two paths
  are homotopic if they admit lifts with the same endpoints in the
  universal covering space.
\end{enumerate}

\Bnn{DECIDING HOMOTOPY AND ISOTOPY}

\begin{enumerate}
\item\emph{Homotopy:} One of the first and most studied problems regarding
  curves on surfaces is concerned with homotopy tests: (1) The
  \emph{contractibility problem}:\tindex{contractibility problem}
  Is a given closed curve (or, equivalently
  here, loop) contractible?
  (2) The \emph{free homotopy problem}:\tindex{free homotopy!problem}
  Are two given closed curves (freely) homotopic?  These problems translate to
  central problems from group theory, in the special case of fundamental
  groups of surfaces: Given a finitely generated group, presented in the
  form of generators and relations, (1) does a given word in the generators
  represent the trivial element of the group (the \emph{word problem}\tindex{word problem})?
  Do two given words in the generators represent conjugate elements in the
  group (the \emph{conjugacy problem}\tindex{conjugacy problem})?

  In computational geometry, these problems are studied in the following
  context: The input is a cellularly embedded graph~$G$ and one or two
  closed curves in~$G$, represented as closed walks in~$G$.  There exist
  linear-time (and thus optimal) algorithms for both the contractibility
  and the free homotopy problems~\cite{lr-hts-12,ew-tcsr-13}.  (An earlier
  article~\cite{dg-tcs-99} claims the same results, but it is
  reported~\cite{lr-hts-12} that the algorithm for free homotopy in that
  article has a subtle flaw.)  The approaches rely on the construction of a
  part of the universal covering space, or on results from small
  cancellation theory in group theory~\cite{gs-sctag-90}.  We remark that
  Dehn's algorithm~\cite{d-tkzf-12} can be implemented in linear time, but
  assuming that the surface is fixed and that the graph has a single face,
  which the other algorithms mentioned above do not require.

\item\emph{Isotopy:} Deciding whether two simple closed curves are isotopic
  can also be done in linear time, because this equivalence relation is a
  simple refinement of homotopy for simple closed curves~\cite{e-c2mi-66}.
  Deciding isotopy of graph embeddings is more complicated, but can also be
  done efficiently, since it essentially reduces to homotopy
  tests for closed curves~\cite{cm-tgis-14}.

\item\emph{Minimum-cost homotopies:} Often, when it is known that two
  curves are homotopic, one would like to compute a ``reasonable''
  homotopy.  Relevant questions include finding a homotopy that sweeps the
  minimum possible area (in a discretized sense)~\cite{cw-msbc2-13}, or has
  the minimum possible number of ``steps''; a homotopy in which the maximum
  length of the intermediate curves is minimal (``height'' of the
  homotopy)~\cite{cl-hh-09}; a homotopy in which the maximum distance
  traveled by a point from the first to the second curve is minimal
  (``width'' of the homotopy---this is related to the \emph{homotopic
    Fr\'echet distance})~\cite{hnss-hwydm-16}; etc.  Several of these
  questions have been studied only in the case of the plane, and extensions
  to surfaces are still open.
\end{enumerate}

\Bnn{ELEMENTARY MOVES AND UNCROSSING}

\begin{figure}[h]
  \centering
  \includegraphics[width=\linewidth]{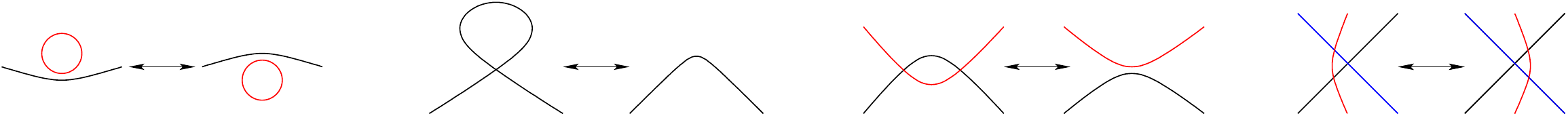}
  \caption{The four Reidemeister moves, up to ambient isotopy.  The
    pictures represent the intersection of the union of the curves with a
    small disk on~$\surf$; in particular, in these pictures, the regions
    bounded by the curves are homeomorphic to disks, and no other parts of
    curves intersect the parts of the curves shown.}
  \label{F:reidemeister}
\end{figure}

\begin{enumerate}
\item\emph{Elementary moves:} Every family of closed curves in general
  position can be made minimally crossing by a finite sequence of
  Reidemeister moves\tindex{Reidemeister moves}, described in Figure~\ref{F:reidemeister}.  If a
  closed curve has $k$ self-crossings, $\Omega(k^2)$ Reidemeister moves can
  be needed; this is tight if the curve is homotopic to a simple curve, but
  in general no subexponential upper bound seems to be
  known~\cite{ce-upc-16}.  Actually, one can deform a family of curves
  continuously to make it minimally crossing without increasing the total
  number of crossings at any step, and moreover, in a minimally crossing
  family, each curve is itself minimally self-crossing, and each pair of
  curves is minimally crossing~\cite{gs-mcmcr-97} (see also
  \cite{hs-scs-94}).  There are other characterizations of curves not in
  minimally crossing position~\cite{hs-ics-85}.
\item\emph{Making curves simple:} Let $G$ be a graph cellularly embedded on
  a surface~$\surf$.  One can decide whether an input curve, represented by
  a closed walk in~$G$, is homotopic to a simple closed curve in~$\surf$ in
  near-linear time.  More generally, one can compute the minimum number of
  self-intersections of a curve in~$\surf$ homotopic to an input closed
  walk in~$G$, and the minimum number of intersections between two curves
  in~$\surf$ respectively homotopic to two input closed walks in~$G$, in
  quadratic time~\cite{dl-cginc-17}.
\item\emph{Untangling curves by a homeomorphism:}\tindex{untangling curves}
  Given two families of
  disjoint, simple curves, one can try to minimize the number of crossings
  between them by changing one of them by a homeomorphism of the surface;
  some bounds are known on the number of crossings that one can
  achieve~\cite{mstw-utsnc-13}.
\item\emph{Simultaneous graph drawing:}\tindex{graph drawing!simultaneous}
  This also relates to the problem of
  embedding two input graphs on the same surface in a way that the
  embeddings cross each other few times.  Here also some results are
  known~\cite{n-cngep-01,rs-tmlrs-05,hkmt-spegs-16}; one can also require
  both combinatorial maps to be fixed.
\item\emph{Number of homotopy classes:} How many simple closed curves in
  different homotopy classes can one draw such that they pairwise cross at
  most~$k$ times, for a given integer~$k$?  On orientable surfaces of genus
  $g\ge2$ without boundary and $k=0$, the answer is $3g-2$ (a pants
  decomposition, see below, together with a contractible closed curve).
  The problem is more interesting for larger values of~$k$; it was recently
  proved that, for fixed~$k$, the number of curves one can draw is
  polynomial in the genus~\cite{p-aimo-15}.
\end{enumerate}

\A{OPTIMIZATION: SHORTEST CURVES AND GRAPHS}\label{sec:optimalization}
\noindent
The problem of computing shortest curves and graphs satisfying certain
topological properties on surfaces has been widely considered.  This leads
to problems with a flavor of combinatorial optimization.

For these problems to be meaningful, a metric must be provided.  In
computational geometry, one could naturally consider piecewise linear
surfaces in some Euclidean space (perhaps~$\R^3$); however, efficient
algorithms for computing shortest paths in such
surfaces~\cite{mmp-dgp-87,ch-spp-96} need additional assumptions because
distances involve square roots, which leads to deep and unrelated questions
on the complexity of comparing sums of square roots~\cite{b-csrpt-91}.
Furthermore, in the context of graph problems in the specific case of
surface-embedded graphs (Section~\ref{sec:other-models} below),
that model would be insufficient.  The notions of combinatorial and
cross-metric surfaces, defined below, have been developed to avoid these
technical distractions, and are suitable in various settings.  On the other
hand, with an oracle for shortest path computations, several of the results
in this section extend to more geometric settings, for example piecewise
linear surfaces in some Euclidean space (see, e.g.,
\cite[Sect.~3.6]{ew-gohhg-05}).

\Bnn{GLOSSARY}

\C{Discrete metrics on surfaces}


\begin{gllist}
\item {\trmbitx Combinatorial surface:}\tindex{combinatorial!surface}\tindex{surface!combinatorial}\quad
  A combinatorial surface is the
  data of a cellular graph embedding~$G$, with positive weights on the
  edges.  The only allowed curves are walks in~$G$; the length of a curve
  is the sum of the weights of the edges of~$G$ traversed by the curve,
  counted with multiplicity.  Algorithmically, curves are stored as closed
  walks in~$G$.  The complexity of the combinatorial surface is the
  complexity of the embedding~$G$ (asymptotically, its number of edges).
\item {\trmbitx Cross-metric surface:}\tindex{cross-metric surface}\tindex{surface!cross-metric}\quad
  A cross-metric
  surface~\cite{ce-tnpcs-10} is also the data of a cellular graph
  embedding~$G$ on some surface~$\surf$, with positive weights on the
  edges.  However, in contrast to the combinatorial surface model, here the
  curves are drawn on the surface~$\surf$ in \emph{general position} with
  respect to~$G$; the length of a curve is the sum of the weights of the
  edges of~$G$ crossed by the curve, counted with multiplicity.
  Algorithmically, a family of curves (or a graph) on a cross-metric
  surface is stored by the combinatorial map of the arrangement of that
  family of curves (or graph) together with~$G$.  The complexity of the
  cross-metric surface is the complexity of the embedding~$G$
  (asymptotically, its number of edges).

  Without loss of generality, one could draw the curves in a neighborhood
  of the dual graph~$G^*$ of~$G$.  Pushing them completely onto~$G^*$ would
  transform them into curves on the combinatorial surface defined by~$G^*$.
  However, the cross-metric surface defined by~$G$ retains more information
  than the combinatorial surface defined by~$G^*$: In the latter case, when
  curves share edges of~$G^*$, they automatically overlap; the cross-metric
  model allows us to make them disjoint except at some well-defined crossing
  points.  (We should point out that it is still possible to define the
  notion of crossing between two curves in a combinatorial surface, but
  this is still insufficient for some of the algorithms described below.)
\end{gllist}

\begin{figure}[h]
  \centering
  \pf{d}\pf{n}\pf{s}
  \includegraphics[width=\linewidth]{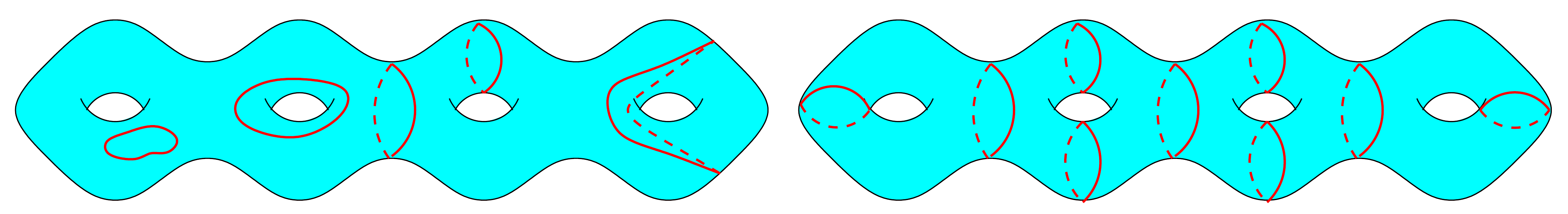}
  \caption{Left: Some closed curves on surfaces, (d) disk-bounding, (n)
    non-separating, (s) splitting.  Right: A pants decomposition of a
    surface.}
  \label{F:curves}
\end{figure}

\C{Types of simple closed curves}
\smallskip
\noindent
Let $\gamma$ be a simple closed curve in the interior of a
surface~$\surf$.  See Figure~\ref{F:curves}.
\begin{gllist}
\item {\trmbitx Disk-bounding curve:}\tindex{curve!disk-bounding}\quad
$\gamma$ is disk-bounding if the
  surface $\surf$ cut along~$\gamma$ (denoted by~$\surf\cut\gamma$) has two
  connected components, one of which is homeomorphic to the disk.
\item {\trmbitx Separating curve:}\tindex{curve!separating}\quad $\gamma$ is separating if
  $\surf\cut\gamma$ has two connected components.
\item {\trmbitx Splitting curve:}\tindex{curve!splitting}\quad $\gamma$ is splitting if $\gamma$ is
  separating but not disk-bounding.
\item {\trmbitx Essential curve:}\tindex{curve!essential}\quad $\gamma$ is essential if no
  component of $\surf\cut\gamma$ is a disk or an annulus.
\end{gllist}

\C{Topological decompositions}

\begin{gllist}
\item {\trmbitx Cut graph:}\tindex{cut graph}\quad A cut graph is a graph~$G$ embedded on a
  surface~$\surf$ such that $\surf\cut G$ is homeomorphic to a closed disk.
\item {\trmbitx System of loops:}\tindex{loop!system of}\tindex{system of loops}\quad
  A system of loops on a surface
  without boundary is a cut graph with a single vertex.  See
  Figure~\ref{F:schema}.
\item {\trmbitx Canonical system of loops:}\tindex{system of loops!canonical}\quad
  A system of loops~$G$ on a
  surface without boundary~$\surf$ is canonical if the edges of the
  polygon~$\surf\cut G$ appear in the same order as in a canonical
  polygonal schema (see Section~\ref{sec:surfaces})
\item {\trmbitx Pants decomposition:}\tindex{pants decomposition}\tindex{decomposition!pants}\quad
  A pants decomposition of an
  orientable surface~$\surf$ is a family~$\Gamma$ of simple, disjoint
  closed curves on~$\surf$ such that $\surf\cut\Gamma$ is a disjoint union
  of pairs of pants.  See Figure~\ref{F:curves}.
\item {\trmbitx Octagonal decomposition:}\tindex{octagonal decomposition}\tindex{decomposition!octagonal}\quad
  An octagonal decomposition
  of an orientable surface~$\surf$ without boundary is a family~$\Gamma$ of
  closed curves on~$\surf$ such that each (self-)intersection point
  in~$\Gamma$ is a crossing between exactly two closed curves, and each
  face of the arrangement of~$\Gamma$ on~$\surf$ is an octagon (a disk with
  eight sides).
\end{gllist}


\C{Homology}
\noindent
In the context of graphs on surfaces, \emph{one-dimensional homology on
  surfaces over the field $\Z/2\Z$} is used; it can be described somewhat
more concisely than more general homology theories.  Let $\surf$ be a
surface.  Here we assume graph embeddings to be piecewise linear (with
respect to a fixed triangulation of~$\surf$).

\begin{gllist}
\item {\trmbitx Homological sum:}\tindex{homological sum}\quad By the previous assumption, the
  closure of the symmetric difference of the images of two graph embeddings
  $G$ and~$G'$ is the image of some graph embedding~$G''$, called the
  homological sum of $G$ and~$G'$.  ($G''$ is defined up to subdivision of
  edges with degree-two vertices, insertion of isolated vertices, and the
  reverse operations; here, graph embeddings are considered up to such
  operations.)
\item {\trmbitx Homology cycle:}\tindex{homology cycle}\tindex{cycle!homology}\quad
  A graph~$G$ embedded on~$\surf$ is a
  homology cycle if every vertex of~$G$ has even degree.  The set of
  homology cycles forms a vector space over the field $\Z/2\Z$: The empty
  graph is the trivial element and addition is the homological sum.
\item {\trmbitx Homology boundary:}\tindex{homology boundary}\tindex{boundary!homology}\quad
  A graph~$G$ embedded on~$\surf$ is
  a homology boundary if the faces of~$G$ can be colored in two colors, say
  black and white, such that $G$ is the ``boundary'' between the two
  colors: Exactly one side of each edge of~$G$ is incident to a black face.
  The set of homology boundaries forms a vector space over $\Z/2\Z$.  Every
  homology boundary is a homology cycle.
\item {\trmbitx Homology group:}\tindex{homology group}\tindex{group!homology}\quad
  It is the $\Z/2\Z$-vector space,
  denoted by~$H_1(\surf)$, that is the quotient of the homology cycles by
  the homology boundaries.  A graph embedding is {\trmbitx homologically
    trivial}\tindex{homology!trivial} if it is a homology boundary.
\end{gllist}
The homology of sets of loops or closed curves can be defined similarly,
because these loops and closed curves are the images of some graph
embedding.  Using the more advanced theory of \emph{singular homology}\tindex{homology!singular}
one can remove the restriction of dealing with piecewise-linear graph embeddings.

\Bnn{BASIC PROPERTIES}
\begin{enumerate}
\item A simple closed curve is disk-bounding if and only if it is
  contractible.
\item A simple closed curve is separating if and only if it is
  homologically trivial.
\item The homology group of a surface~$\surf$ without boundary has
  dimension~$\bar{g}$, the Euler genus of~$\surf$, and is generated by the
  loops appearing on the boundary of a canonical polygonal schema.
\end{enumerate}

\Bnn{SHORTEST CURVES}
\noindent
Deciding whether a simple closed curve in a cross-metric (or combinatorial)
surface is separating or disk-bounding can be done in time linear in the
size of the data structure used to store the cellular graph and the curve;
this boils down to determining whether some graph is connected, or whether
some surface is a disk (which is easy using Euler's formula).  Here we
consider the optimization version, by looking for shortest curves with a
given topological type in a combinatorial or cross-metric surface.
Non-disk-bounding or non-separating curves are of particular interest,
because cutting along such a curve simplifies the topology of a surface.
Below we use \emph{non-trivial} as a shorthand for either non-disk-bounding
or non-separating.

\begin{table}[htb]
  {\begin{tabular}{|r|l|l|}
   \hline
\rule[-4pt]{0pt}{13pt} & {UNDIRECTED} & {DIRECTED} \\\hline
  \pbox[t]{.3\linewidth}{\raisebox{-2.1cm}{\rotatebox{45}{WEIGHTED}}} &%
  \pbox[t]{.6\linewidth}{%
    $\vphantom{e^{e^{e^e}}}\bm{O(n^2\log n)}$~\cite{eh-ocsd-04}\\
    {\tabcolsep 0pt\partabular.{
      \}}{ll}{$O(g^{3/2}n^{3/2}\log n)$\ \ & non-sep\\
      $g^{O(g)}n^{3/2}$ & non-db}}
    \cite{cm-fsnnc-07}
    \\
    $g^{O(g)}n\log n$~\cite{k-csntc-06}\\
    $O(g^3n\log n)$~\cite{cc-msspg-07}\\
    $\bm{O(g^2n\log n)}$~\cite{cce-msspe-13}\\
    $g^{O(g)}n\log\log n$~\cite{insw-iamcm-11}\\
    $\bm{{2^{O(g)}n\log\log n}}$~\cite{f-sntcd-13}\\
    ${\bm{O(gn\log n)}}$  for 2-approx.~\cite{eh-ocsd-04}$\vphantom{e_{e_{e_e}}}$
     } &
  \pbox[t]{.6\linewidth}{%
    $O(n^2\log n)$~\cite{ccl-fsntc-16}\\
    $\bm{O(g^{1/2}n^{3/2}\log n)}$~\cite{ccl-fsntc-16}\\
    $2^{O(g)}n\log n$ non-sep~\cite{en-mcsnc-11}\\
    {\tabcolsep 0pt\partabular.{
        \}}{ll}{$\bm{O(g^2n\log n)}$\ \ & non-sep\\
    $g^{O(g)}n\log n$ & non-db}}~\cite{e-sntcd-11}\\
    $\bm{O(g^3n\log n)}$ non-db~\cite{f-sntcd-13}}
    \\\hline
  \pbox[t]{.3\linewidth}{\raisebox{-1.3cm}{\rotatebox{45}{UNWEIGHTED}}} &%
  \pbox[t]{.6\linewidth}{%
    $\vphantom{e^{e^{e^e}}}$$O(n^3)$~\cite{t-egsnc-90} (see~\cite{mt-gs-01})\\
    $\bm{O(n^2)}$~\cite{ccl-aeweg-12}\\
    $\bm{O(gnk)}$~\cite{ccl-aeweg-12}\\
    ${\bm{O(gn/\varepsilon)}}$  for $(1+\varepsilon)$-approx.~\cite{ccl-aeweg-12}\
     \ \
}
    &
  \pbox[t]{.6\linewidth}{
      $\bm{O(n^2)}$~\cite{ccl-fsntc-16}\\
      $\bm{O(gnk)}$~\cite{ccl-fsntc-16}
}\\
\hline
\end{tabular}}
  \caption{Algorithms for shortest non-trivial closed curves on surfaces
    without boundary, depending on whether the graph is weighted and whether it is
    directed.  ``Non-sep'' and ``non-db'' mean non-separating and
    non-disk-bounding, respectively; $k$ is the size of the output.  The best
    complexities known to date are in bold (there can be several of them in
    each category due to the tradeoff between $g$, $n$, and~$k$).  Of
    course, the undirected case reduces to the directed case, and the
    unweighted case reduces to the weighted case; in each cell, we do not
    repeat the algorithms that are available for more general scenarios.}
   \label{t:nontrivial}%
\end{table}

\begin{enumerate}
\item\emph{Structural properties:}\tindex{loop!shortest}\tindex{shortest loop}
  In a combinatorial surface, a shortest
  noncontractible or non-null-homologous loop based at a vertex~$x$ is
  made of two shortest paths from~$x$ and of a single edge (this is the
  so-called 3-path condition~\cite{t-egsnc-90}).  It follows that the
  globally shortest non-contractible and non-null-homologous closed curves
  do not repeat vertices and edges, and are also shortest non-disk-bounding
  and non-separating closed curves.  More generally, in the algorithms
  mentioned below, a typical tool is to prove a bound on the number of
  crossings between the (unknown) shortest curve and any shortest path.
\item\emph{Different scenarios for shortest non-trivial curves:}\tindex{curve!shortest}\tindex{shortest curve}
  Table~\ref{t:nontrivial}
  summarizes the running times of the known algorithms.  In such problems,
  it is relevant to look for more efficient algorithms in the case where
  the genus~$g$ is smaller compared to the complexity~$n$ of the graph
  defining the surface.  The standard scenario, which is the only one
  considered elsewhere in this chapter, is that of a combinatorial (or
  equivalently, cross-metric) surface (the undirected, weighted case, in
  the upper left corner in Table~\thechapter.5.1
  ).  One can also aim for faster algorithms in the \emph{unweighted} case
  (unit weights).  Finally, one can extend the techniques to the case of
  \emph{directed} graphs, where the edges of the combinatorial surface are
  directed and can only be used in a specified direction (equivalently, the
  edges of the cross-metric surface can only be crossed in a specific
  direction).
\item\emph{Other topological types:}\tindex{path!shortest}\tindex{shortest path}
  Shortest simple closed curves of other
  topological types have been investigated as well (in the following, $n$
  denotes the complexity of the cross-metric surface): shortest
  splitting curves~\cite{ccelw-scsh-08} (NP-hard, but computable in
  $O(n\log n)$ time for fixed genus); shortest essential
  curves~\cite{ew-csec-10} ($O(n^2\log n)$ time, or $O(n\log n)$ for fixed genus
  and number of boundaries---in this case, surfaces with boundary require
  more sophisticated techniques); and non-separating curves which are
  shortest in their (unspecified) homotopy class~\cite{cdem-fotc-10}
  ($O(n\log n)$).
\item\emph{Shortest homotopic curves:} A slightly different problem is that
  of computing a shortest curve homotopic to a given curve (either a path
  or a closed curve); this is also doable in small polynomial time, using
  octagonal decompositions to build a part of the universal covering
  space~\cite{ce-tnpcs-10} (earlier algorithms dealt with simple curves
  only, with an iterated shortening process that leads to a global optimum~\cite{cl-oslos-05,cl-opdsh-07}).
\item\emph{Shortest paths:} All these algorithms rely on shortest path
  computations on combinatorial (or cross-metric) surfaces, which can be
  done in $O(n\log n)$ time using Dijkstra's algorithm~\cite{d-ntpcg-59}
  classically speeded up with Fibonacci heaps~\cite{ft-fhuin-87} in the
  primal (or dual) graph.  This actually computes the shortest paths from a
  single source to all other vertices of the combinatorial surface.  Other
  algorithms are available for computing multiple shortest paths quickly
  under some conditions on the locations of the
  endpoints~\cite{cce-msspe-13}.
\end{enumerate}

\Bnn{SHORTEST DECOMPOSITIONS}
\noindent
Decompositions of surfaces are central in topology; for example, the
standard proof of the classification theorem transforms an arbitrary cut
graph into a canonical system of loops.  Many algorithms described in the
previous subsection rely on topological decompositions and their
properties.
\begin{enumerate}
\item\emph{Shortest cut graph:} The problem of computing a shortest cut
  graph on a cross-metric surface has been extensively studied.  Computing
  the shortest cut graph is NP-hard, but there is an
  $O(\log^2g)$-approximation algorithm that runs in $O(g^2n\log n)$
  time~\cite{eh-ocsd-04}.  Moreover, for every $\varepsilon>0$ one can
  compute a $(1+\varepsilon)$-approximation in $f(\varepsilon,g)\cdot n^3$
  time, for some function~$f$~\cite{cm-fptas-15}.  If one is looking for a
  shortest cut graph with a specified vertex set~$P$ (for example, a
  shortest system of loops with given basepoint~\cite{ew-gohhg-05}), then
  there is an algorithm with running time $O(n\log
  n+gn+|P|)$~\cite{c-scgsp-10}.  At the root of several of these articles
  lies the \emph{tree-cotree property}~\cite{e-dgteg-03}: If $G$ is a
  cellular graph embedding, there exists a partition $(T,C,X)$ of the edges
  of~$G$ such that $T$ is a spanning tree of~$G$ and the edges dual to~$C$
  form a spanning tree of the dual graph~$G^*$.  Contracting~$T$ and
  deleting~$C$ transforms~$G$ into a system of loops, each loop
  corresponding to an element of~$X$.
\item\emph{Other topological decompositions:} Some canonical system of
  loops (for orientable surfaces without boundary) can be computed in
  $O(gn)$ time~\cite{lpvv-ccpso-01}.  An octagonal decomposition or a pants
  decomposition made of closed curves which are as short as possible in
  their respective homotopy classes can be computed in $O(gn\log n)$
  time~\cite{ce-tnpcs-10}.  But in general the complexity of computing
  shortest such decompositions is open.  On the other hand, there are
  bounds on the maximum length of some decompositions, assuming that the
  combinatorial surface is an unweighted triangulation, or, dually, that
  the cross-metric surface is unweighted and each vertex has degree
  three~\cite{chm-dsidt-15}.
\item\emph{Stretch:} Let $\surf$ be a cross-metric surface, and let $G$ be
  the associated embedded graph.  The stretch of~$\surf$ is the minimum of
  the \emph{product} of the lengths of~$\gamma$ and~$\delta$, over all
  closed curves $\gamma$ and~$\delta$ crossing exactly once.  This quantity
  is related to the planar crossing number and the size of a largest
  toroidal grid minor of~$G^*$~\cite{hc-acnge-10}, and can be computed in
  small polynomial time~\cite{cch-cseg-14}.
\end{enumerate}

\Bnn{HOMOLOGY AND ITS RELATION TO CUTS AND FLOWS}
\noindent
As hinted above, homology is useful because a simple closed curve is
separating if and only if it is null-homologous; the algorithms for
computing shortest non-separating closed curves actually compute shortest
non-null-homologous closed curves, which turn out to be simple.

Homology is a natural concept; in particular, it is interesting to look for
a family of closed curves, of minimum total length, the homology classes of
which generate the homology group.  Some efficient algorithms have been
given for this purpose~\cite{ew-gohhg-05}, also in connection with an
algorithm to compute a minimum cycle basis of a surface-embedded
graph~\cite{bcfn-mchbs-16}.

Another reason for the importance of homology is its relation to cuts:
Given a graph~$G$ cellularly embedded on a surface~$\surf$ without
boundary, the $(s,t)$-cuts in~$G$ are dual to the subgraphs of~$G^*$ in
some fixed homology class on the surface obtained from~$\surf$ by removing
the faces of~$G^*$ containing $s$ and~$t$. Thus, computing minimum cuts
amounts to computing shortest homologous subgraphs.  This property has been
exploited to study general graph problems, where better algorithms can be
designed in the specific case of graphs embedded on a fixed surface, to:
\begin{enumerate}
\item compute minimum $(s,t)$-cuts\tindex{st-cut@$(s,t)$-cut} in near-linear
  time~\cite{cen-mcshc-09,en-mcsnc-11}.  The best algorithm runs in
  $2^{O(g)}n\log n$ time, where $g$ is the genus~\cite{en-mcsnc-11}, and
  relies on the \emph{homology cover}\tindex{homology cover}, a particular type of covering space;
\item compute maximum $(s,t)$-flows faster, by exploiting further the
  duality between flows and cuts~\cite{cen-hfcc-12,benw-apmcn-16};
\item count and sample minimum $(s,t)$-cuts
  efficiently~\cite{cfn-csmcg-14};
\item compute global minimum cuts efficiently (without fixing $s$
  and~$t$)~\cite{efn-gmcse-12};
\item deal with other problems, e.g., to compute the edge expansion and
  other connectivity measures~\cite{p-deeoc-13} or to bound the space
  complexity of bipartite matching~\cite{dgkt-ibbms-12}.
\end{enumerate}

\A{ALGORITHMS FOR GRAPHS EMBEDDED ON A FIXED SURFACE}\label{S:algo}
\noindent
Some general graph problems can be solved faster in the special case of
graphs embedded on a fixed surface.  Examples include cut and flow problems
(see previous section), multicommodity problems, domination and
independence problems, connectivity problems (Steiner tree, traveling
salesman problem, etc.), disjoint paths problems, shortest paths problems,
subgraph problems, and more.

Sometimes the problems are solvable in polynomial-time on arbitrary graphs,
and the goal is to obtain faster algorithms for surface-embedded graphs.
But in many cases, the problems considered are NP-hard on arbitrary graphs,
and polynomial-time algorithms are obtained for graphs embeddable on a
fixed surface (occasionally by fixing some other parameters of the
problem).  Typically, optimization problems are considered, in which case
it is relevant to look for approximation algorithms.

The methods involved usually combine topological aspects (as described
above) with techniques from structural and algorithmic graph theory.

\Bnn{GLOSSARY}

\begin{gllist}
\item {\trmbitx Minor:}\tindex{minor!graph}\quad A graph~$H$ is a minor of another graph~$G$ if
  $H$ can be obtained from~$G$ by removing edges and isolated vertices, and
  contracting edges.
\item {\trmbitx Minor-closed family:}\tindex{graph!minor-closed family of}\tindex{minor-closed family!graph}\quad
  A family~$\FF$ of graphs is
  minor-closed if every minor of a graph in~$\FF$ is also in~$\FF$.
\item {\trmbitx Tree decomposition:}\tindex{tree!decomposition}\tindex{decomposition!tree}\quad
 A tree decomposition of a
  graph~$G=(V,E)$ is a tree~$T$ in which each node is labeled by a subset
  of~$V$, such that:
  \begin{itemize}
  \item\ for each $v\in V$, the set of nodes in~$T$ whose labels contain~$v$
    induces a non-empty connected subtree of~$T$, and
  \item\ if $G$ has an edge connecting vertices $u$ and~$v$, then the label
    of at least one node of~$T$ contains both $u$ and~$v$.
  \end{itemize}
\item {\trmbitx Width:}\tindex{width!graph}\quad The width of a tree decomposition is the
  maximum cardinality of the labels minus one.
\item {\trmbitx Treewidth:}\tindex{treewidth}\quad The treewidth of a graph~$G$ is the
  minimum width of a tree decomposition of~$G$.
\end{gllist}

\Bnn{SURVEY OF TECHNIQUES}
\noindent
Central to algorithmic and structural graph theory is the study of
minor-closed families of graphs; by a deep result of Robertson and
Seymour~\cite{rs-gm20w-04}, for each such family~$\FF$, there is a
\emph{finite} set~$X_\FF$ of graphs such that $G\in\FF$ if and only if no
graph in~$X_\FF$ is a minor of~$G$.  We refer to~\cite{km-srpag-07} for a
survey on these structural aspects.

The graphs embeddable on a fixed surface form a minor-closed family, and
have the benefit that they can be studied using topological techniques.
Robertson and Seymour provide a decomposition theorem for minor-closed
families of graphs involving graphs embeddable on a fixed
surface~\cite{rs-gm16e-03}; efficient algorithms for surface-embedded
graphs are sometimes extended to minor-closed families of graphs (different
from the family of all graphs).

It is impossible to list all results in algorithms for surface-embedded
graphs here, so we focus on general methods.  Several algorithms are based
on topological techniques described in the previous sections (in
particular, shortest non-trivial curves or shortest decompositions), in
several cases with advanced algorithmic
techniques~\cite{en-crpse-11,kks-lsado-11,es-noaaa-14,ppsl-nsspp-14}.
Sometimes the same techniques have led to new results for planar
graphs~\cite{e-mfpsp-10,en-sncwp-11,c-mpbgg-15}.  Methods applicable to
several algorithmic problems have also emerged, in many cases extending
previous ones invented for planar graphs:
\begin{enumerate}
\item\emph{Graph separators and treewidth:} Let $G$ be a graph with
  $n$~vertices embedded on a surface with genus~$g$.  In linear time, one
  can compute a balanced separator of size $O(\sqrt{gn})$, namely, a set of
  $O(\sqrt{gn})$ vertices whose removal leaves a graph without connected
  component of more than $2n/3$ vertices~\cite{ght-stgbg-84,e-dgteg-03}.
  Also, the treewidth of~$G$ is $O(\sqrt{gn})$.
\item\emph{Dynamic programming:} Small treewidth implies efficient
  algorithms using dynamic programming in arbitrary graphs.  When the graph
  is embedded, one can exploit this fact to obtain algorithms with smaller
  dependence on the treewidth for some
  problems~\cite{b-ssdsi-12,rst-aenps-13,rst-dpgs-14}.
\item\emph{Irrelevant vertex technique:} Several graph problems enjoy the
  following property~\cite{t-gmpad-12}: If the input graph has large
  treewidth, there exists an irrelevant vertex, whose removal creates an
  equivalent instance of the problem (e.g., a vertex at the center of a
  large grid minor).  This property is widely used in structural graph
  theory and has been exploited several times in the context of algorithms
  for surface-embedded graphs~\cite{kr-ccnlt-07,kt-ccgs-12,rs-gm12i-12}.
\item\emph{Polynomial-time approximation schemes (PTASs):}
  Baker~\cite{b-aancp-94} has introduced a technique for designing
  approximation schemes for some optimization problems with local
  constraints in planar graphs: She has showed that one can delete a small
  part of the input graph without changing too much the value of the
  solution and such that the resulting graph has small treewidth.  The
  technique has been extended to graphs embeddable on a fixed
  surface~\cite{e-dtmcg-00}, to graphs that can be drawn on a fixed surface
  with a bounded number of crossings per edge~\cite{gb-agefc-07}, and to
  more general \emph{contraction-closed} problems where contraction instead
  of deletion must be used~\cite{k-ltasp-05,dhm-aacd-10}.  A crucial step
  in making the latter technique effective is the construction of a
  \emph{spanner}: In the case of a minimization problem, this is a subgraph
  of the input graph containing a near-optimal solution and whose weight is
  linear in that of the optimal solution.  \emph{Brick decomposition} is a
  technique that builds spanners for some problems, originally in planar
  graphs, but also sometimes in graphs on surfaces~\cite{bdt-ptass-14}.
\item\emph{Bidimensionality:} This
  theory~\cite{t-bpa-15,dfht-spabg-05,dht-btbgg-06,dh-btaa-08} applies to
  minimization problems on unweighted graphs where contracting an edge of
  the graph does not increase the value of the solution, and where the
  value of the solution in grid graphs (and generalizations) is large.  It
  leads to output-sensitive algorithms for graphs embeddable on a fixed
  surface with running time of the form $2^{O(\sqrt{k})}\cdot n^{O(1)}$,
  where $k$ is the value of the solution and $n$ is the input size. This
  also provides PTASs in some cases~\cite{dh-bncbf-05}.  For the problems
  where bidimensionality applies, PTASs can sometimes also be obtained in
  weighted graphs using a different framework~\cite{cckmm-acdwb-16}.
\item\emph{Stochastic embeddings:} Let $G=(V,E)$ and $G'=(V',E')$ be
  positively edge-weighted graphs.  A \emph{non-contracting metric
    embedding}~$f$ from $G$ to~$G'$ is a mapping from $V$ to~$V'$ such that
  $d'(f(x),f(y))\ge d(x,y)$ for each $x,y\in V$, where $d$ and~$d'$
  represent shortest path distances in $G$ and~$G'$, respectively.  The
  \emph{distortion} of~$f$ is the maximum of $d'(f(x),f(y))/d(x,y)$ over
  all $x\ne y\in V$ (see Chapter~\chapIndykA).  Every graph~$G$
  embeddable on an orientable
  surface~$\surf$ of genus~$g$ admits a probability distribution of
  non-contracting metric embeddings into planar graphs such that for each
  $x,y\in V$, one has $\E[d'(f(x),f(y))]\le O(\log g)\cdot d(x,y)$, where
  the expectation is over all $f$ in the distribution~\cite{s-osp-10}.
  This reduces several optimization problems on graphs on~$\surf$ to the
  same problem in planar graphs, up to the loss of an $O(\log g)$ factor.
  Actually, such a distribution can be computed in polynomial time even if
  no embedding of~$G$ on~$\surf$ is known~\cite{ms-pus-12}.
\end{enumerate}

\A{OTHER MODELS}\label{sec:other-models}
\noindent
A rather large number of results relate to the concepts described in this
chapter, and it would be impossible to cover them all.  Below, we provide a
selection of miscellaneous results that consider other models for
representing graphs on surfaces.

\Bnn{COMPUTATIONAL TOPOLOGY IN THE PLANE WITH OBSTACLES}
\noindent
The plane minus finitely many points or polygons (``obstacles'') forms a
(non-compact) surface~$\surf$.  Taking any cellular graph embedding
on~$\surf$ makes~$\surf$ a combinatorial (or cross-metric) surface, so most
of the topological algorithms above apply.  However, it is much more
natural to consider arbitrary piecewise-linear curves in~$\surf$, whose
length is defined by the Euclidean metric.  In this model, $\surf$ is
defined by the obstacles (a finite set of disjoint simple polygons, for
simplicity of exposition); curves are arbitrary polygonal lines avoiding
the interior of the obstacles.  Some of the problems defined in the
previous sections and related problems have been studied in this model:
\begin{enumerate}
\item \emph{Homotopy and isotopy tests:} There are efficient algorithms to
  test whether two curves are (freely) homotopic~\cite{clms-thpp-04}, or
  whether two graphs are isotopic~\cite{cm-tgis-14}.
\item \emph{Shortest homotopic paths} can be computed efficiently as
  well~\cite{hs-cmlpg-94,b-chspp-03,ekl-chspe-06}; see also Section~\chapShortest.2.
  A variant where several simple and disjoint paths must be shortened while
  preserving their homotopy class and keeping their neighborhoods simple
  and disjoint (i.e., the paths are ``thick'') has also been
  investigated~\cite{gjkmrs-cholr-88}.
\item \emph{Shortest disjoint paths:} Here the goal is to compute disjoint
  paths with minimum total length (or, more precisely, non-crossing paths,
  since in the limit case, the solution may consist of overlapping paths).
  If the endpoints lie on the boundary of a bounded number of obstacles,
  the problem is solvable in polynomial time~\cite{en-sncwp-11}.
\item \emph{Other results} include a constant-factor approximation
  algorithm for the shortest pants decomposition in the case where the
  obstacles are points~\cite{e-stssc-09} and an algorithm for computing the
  homotopic Fr\'echet distance, a measure of similarity between curves that
  takes the obstacles into account topologically~\cite{ccellt-hfdbc-10,hnss-hwydm-16}.
\end{enumerate}

\Bnn{SIMPLE AND DISJOINT CURVES IN GRAPHS}
\noindent
In the cross-metric model, defined by a cellularly embedded graph~$G$, one
can think of curves as being drawn in a neighborhood of~$G^*$.  So,
intuitively, curves are drawn in~$G^*$, but they can share vertices and
edges of~$G^*$ while being simple and pairwise disjoint.

It is very natural, especially in topological graph theory, to forbid such
overlaps: A set of disjoint simple curves cannot repeat any vertex or edge
of~$G^*$.  Many of the problems mentioned in the previous sections make
sense in this setup, which turns out to be generally more difficult to
handle than the cross-metric model.  In this model, the following results
are known (here by \emph{circuit} we mean a closed curve in the graph
without repeated vertex, and containing at least one edge):
\begin{enumerate}
\item Determining whether there exists a separating (resp., splitting)
  circuit is NP-complete~\cite{ccl-fctpe-11}.
\item Determining some contractible (resp., non-contractible, resp.,
  non-separating) circuit, if such a circuit exists, is possible in linear
  time, even if one requires the circuit to pass through a given
  vertex~\cite{ccl-fctpe-11}.
\item Computing a \emph{shortest} contractible circuit is possible in
  polynomial time, but if one requires the circuit to pass through a given
  vertex, the problem becomes NP-hard~\cite{c-fscss-10}.
\item Computing a \emph{shortest} separating circuit is
  NP-hard~\cite{c-fscss-10}.
\item There is a combinatorial characterization on whether curves can be
  made simple and disjoint in the graph by a homotopy on the
  surface~\cite{s-dcphg-91}.  In the case of a planar surface with
  boundaries, this leads to a polynomial-time
  algorithm~\cite[Th.~31]{s-hrm-90}, which in turn has some algorithmic
  consequences on the problem of computing vertex-disjoint paths in planar
  graphs~\cite[Th.~34]{s-hrm-90}.  See also~\cite[Ch.~76]{s-cope-03}.
\end{enumerate}

\Bnn{NORMAL CURVES ON SURFACES}
\noindent
Let $\Gamma$ be a family of disjoint simple closed curves on a
surface~$\surf$ in general position with respect to a triangulation~$T$
of~$\surf$.  A natural way to represent~$\Gamma$, as described in the
previous sections, is by its arrangement with~$T$.
\emph{Normal curves}\tindex{curve!normal}\tindex{normal!curve}
are a more economical representation, at the price of a mild condition: For
every triangle~$t$ of~$T$, the intersection of the image of~$\Gamma$
with~$t$ must be a set of (disjoint simple) paths, called \emph{normal
  arcs}, connecting \emph{different} sides of~$t$.
For such a~$\Gamma$, and for each triangle~$t$ of~$T$, one stores three integers recording the
number of normal arcs connecting each of the three pairs of sides of~$t$.
Overall, $\Gamma$ is described by $3n$~non-negative integers, where $n$~is
the number of triangles in~$T$.  Conversely, given a vector of $3n$
non-negative integers, one can unambiguously reconstruct~$\Gamma$ up to
\emph{normal isotopy}\tindex{isotopy!normal}, that is, up to an ambient isotopy that leaves the
edges of~$T$ globally unchanged.

To store the vector of normal coordinates, $O(n\log(X/n))$ bits are needed,
where $X$ is the number of crossing points of~$\Gamma$ with~$T$.  In
contrast, representing these curves on a cross-metric surface requires at
least to store a constant amount of information per vertex of the
arrangement, which is $\Theta(n+X)$ in total.  So the normal curve
representation can be exponentially compressed compared to the
``na\"\i{}ve'' one.  Despite this, in time polynomial in the input size one
can:
\begin{enumerate}
\item count the number of connected components of a normal curve (note that
  a ``normal curve'' does not have to be connected), and partition these
  components according to their (normal or not) isotopy classes, given by
  their multiplicities and the normal coordinates of a
  representative~\cite{sss-ancs-02,en-tccts-13};
\item decide whether two normal curves are
  isotopic~\cite{sss-ancs-02,en-tccts-13};
\item compute the algebraic~\cite{sss-ancs-02,en-tccts-13} or the
  geometric~\cite{sss-cdtgi-08} intersection number of two normal curves.
  (The algebraic intersection number of $\gamma$ and~$\delta$ is the sum,
  over all crossings between $\gamma$ and~$\delta$, of the sign of the
  crossing, which is $+1$ if $\gamma$ crosses $\delta$ from left to right
  at that crossing point and $-1$ otherwise; this is well-defined if the
  surface is orientable, since it is invariant by isotopy.  The geometric
  intersection number of $\gamma$ and~$\delta$ is the minimum number of
  crossings between curves $\gamma'$ and~$\delta'$ isotopic to $\gamma$
  and~$\delta$.)
\end{enumerate}

These problems have been initially studied using \emph{straight-line
  programs}, a concise encoding of words over a finite alphabet; many
algorithms on words can be solved efficiently using the straight-line
program representation, in particular because straight-line programs can
represent exponentially long words; this leads to efficient algorithms for
normal curves~\cite{sss-ancs-02,sss-cdtgi-08}.  The same and other problems
have been revisited using more topological techniques~\cite{en-tccts-13}.
Normal curves are the lower-dimensional analog of
\emph{normal surfaces}\tindex{surface!normal}\tindex{normal!surface},
widely used in three-dimensional topology.

\A{OTHER RESOURCES}

\noindent
\paragraph{Books.}
Graphs on surfaces from a combinatorial viewpoint are treated in detail
in~\cite{mt-gs-01}; see also~\cite{gt-tgt-87}.  For basic surface topology,
we recommend~\cite{a-bt-83,s-ctcgt-93,h-cit-94}.

\paragraph{Survey.}
\cite{e-cocb-12} surveys optimization problems for surface-embedded graphs,
providing more details on a large fraction of Section~\ref{sec:optimalization}.

\paragraph{Course notes and unpublished material.}
\cite{e-ct-13} provides some notes in computational topology with a strong
emphasis on graphs on surfaces.  \cite{c-tags-12,c-aeg-17} survey some
algorithms for optimization of graphs and curves on surfaces.
\cite{dmst-apgb-11} emphasizes graph algorithms for surface-embedded
graphs.

\Bnn{RELATED CHAPTERS}

\noindent Chapter~\chapPolyhedralMaps: Polyhedral maps 

\noindent Chapter~\chapShortest: Shortest paths and networks 

\noindent Chapter~\chapGraphDraw: Graph drawing 

\Bnn{ACKNOWLEDGMENTS}

Many thanks to Sergio Cabello, Vincent Cohen-Addad, Jeff Erickson, Francis
Lazarus, Arnaud de Mesmay, and Dimitrios Thilikos for their careful reading
of preliminary versions, and for their numerous comments that greatly
improved this chapter.

\small
\newcommand{\etalchar}[1]{$^{#1}$}

\end{document}